# Quantum-aided design for advanced qubit: plasmonium

Feng-Ming Liu[1,2,3], Ming-Cheng Chen[1,2,3], Can Wang[1,2,3], He Chen[1,2,3], Shao-Wei Li[1,2,3], Zhong-Xia Shang[1,2,3], Chong Ying[1,2,3], Jian-Wen Wang[1,2,3], Yong-Heng Huo[1,2,3], Cheng-Zhi Peng[1,2,3], Xiaobo Zhu[1,2,3], Chao-Yang Lu[1,2,3], Jian-Wei Pan[1,2,3]

[1]Hefei National Research Center for Physical Sciences at the Microscale and Department of Modern Physics, University of Science and Technology of China, Hefei, Anhui 230026, China [2]Shanghai Branch, CAS Centre for Excellence and Synergetic Innovation Centre in Quantum Information and Quantum Physics, University of Science and Technology of China, Shanghai 201315, China [3]Shanghai Research Center for Quantum Sciences, Shanghai 201315, China

**Abstract:**

**The increasingly complex quantum electronic circuits with a number of coupled quantum degrees of freedom will become intractable to be simulated on classical computers, and requires quantum computers for an efficient simulation. In turn, it will be a central concept in quantum-aided design for next-generation quantum processors. Here, we demonstrate variational quantum eigensolvers to simulate superconducting quantum circuits with varying parameters covering a plasmon-transition regime, which reveals an advanced post-transmon qubit, "plasonium". We fabricate this new qubit and demonstrate that it exhibits not only high single- and two-qubit gate fidelities (99.85(1)% and 99.58(3)%, respectively), but also a shrinking physical size (by 60%) and a larger anharmonicity (by 50%) than the transmon, which can bring a number of advantages for scaling up multi-qubit devices. Our work opens the way to designing advanced quantum processors using existing quantum computing resources.**

The past half-century has witnessed the rapid development of integrated circuits from several to billions of electronic components. A key enabling method was the electronic computer-aided design software, with which the engineers used the existing computers to calculate and simulate increasingly more complex electronic circuits, to inspire and design next-generation higher-performance processors. Similarly, efficient simulations of the quantum circuits might play an important role in large-scale quantum computers.

The elementary building block of a large-scale quantum computer is an ultra-low-noise qubit, whose performance is crucial for achieving the break-even point in quantum error correction and reducing the resource overhead for fault-tolerant quantum computation[1]. The past decades have witnessed great progress[2-4] in continuously improving the qubit performance. For example, the invention of the transmon qubit[5,6], which has a simple structure and allows high-fidelity quantum gates, has enabled scaling up to high-fidelity processors with 50-70 qubits for demonstrations of quantum computational advantage[7] and surface-code error correction[8]. In the request of ever-growing size and complexity of the quantum computers, designing an advanced post-transmon qubit with an even higher performance is highly desirable.

In principle, advanced noise-protected qubits can be realized at the quantum electronic circuit level by elaborately coupling various quantum degrees of freedom to construct robust quantum states[9-11]. While the simulation of quantum electronic circuits could be done classically on a small scale, generally it is expected that the resources required for the classical simulations grow exponentially as $O(d^N)$ with $N$ coupled $d$-dimensional quantum degrees of freedom. For this purpose, it is therefore interesting to exploit the capability of intermediate-scale quantum computers[12]—for instance, those before full-fledged for prime factoring—to effectively aid the design of future higher-performance quantum computers (Fig. 1a).

In this work, we demonstrate a quantum-aided design of a proof-of-principle model, a modified transmon circuit where an inductance formed by a Josephson-junction array is added in parallel to the original circuit[13] (Fig. 1b). This circuit model has been widely used to establish many successful qubits with different circuit parameters, such as the fluxonium qubit[14], heavy fluxonium[15,16], high-coherence fluxonium[17,18], quasi-charge qubit[19], and quarton qubit[20]. As we will elaborate, the new discovered qubit works in a different parameter regime (Extended Data Fig. 1), is fully compatible with the control technologies as in the widely used transmon, and has remarkable features superior to the transmons.

We note that although this prototypical quantum circuit model can be solved efficiently on classical computers, such as using the recently developed simulation packages such as scqubits[21] and circuitQ[22], our work takes the first step in the direction of quantum-aided design towards multiple degrees of freedom circuits that are beyond the classical simulation method.

In quantum-aided design, one of the core tasks is to solve a series of hard Hamiltonian eigenproblems of parameterized electronic circuits, which is important for identifying the encoding subspace of qubit and calculating relevant physical properties. We start from the quantum computing[23-25] of the energy eigenspectra of the targeted circuit. The system has a non-linear element, Josephson junction of Josephson energy $E_J$, and two linear elements, capacitance $C$ and inductance $L$ of charging and inductive energy $E_C = e^2/2C$ and $E_L = (\hbar/2e)^2/L$, respectively. The system Hamiltonian is

$$H = 4E_C \hat{n}^2 - E_J \cos(\hat{\phi} + \phi_{ext}) + \frac{1}{2} E_L \hat{\phi}^2 \qquad (1)$$

where $\phi$ is the phase across the inductance, and $n$ is the number of cooper pairs tunneled between the two islands of the capacitance which is conjugate to $\phi$. $\phi_{ext}$ is the external flux threading through the loop between the Josephson junction and inductance, in the unit of the reduced superconducting flux quantum $\Phi_0/2\pi = \hbar/2e$. As the quantum variables $n$ and $\phi$ are both continuous, we will use the Fock-state basis of its harmonic part $H_{har} = 4E_C \hat{n}^2 + \frac{1}{2} E_L \hat{\phi}^2$ to discretely encode the system Hamiltonian[26]. Let $a$ and $a^\dagger$ be the annihilation and creation operator, the whole

Hamiltonian becomes

$$H = \omega \hat{a}^\dagger \hat{a} - E_J \cdot \frac{\hat{D}(iA)e^{i\phi_{ext}} + \hat{D}(-iA)e^{-i\phi_{ext}}}{2} \quad (2)$$

where $\omega = \sqrt{8E_C E_L}$ is the transition frequency of $H_{har}$, $A = (2E_C/E_L)^{1/4}$, and $\hat{D}$ is the displacement operator defined as $\hat{D}(\alpha) = \exp(\alpha \hat{a}^\dagger - \alpha^* \hat{a})$. We truncate the Hilbert space to finite dimensions so that it can be mapped to the computational basis of our quantum processor. Here, we keep the Fock states up to 7 photons to provide enough accuracy for our quantum simulation (Fig. 2a).

In this work, we utilized the subspace-search variational quantum eigensolver to find the lowest three energy eigenstates simultaneously[27]. We initialize the quantum state in the lowest three Fock states and then execute the same unitary quantum circuit. As the input states are orthogonal, the final states after the circuit are also orthogonal. We define the cost function for variational optimization of the quantum circuit as

$$F_{cost} = 5 \times E_0 + 4 \times E_1 + 2 \times E_2 \quad (3)$$

where the weighting coefficients for lower eigenstates are larger than higher eigenstates, and $E_0$, $E_1$ and $E_2$ are the measured expectation values of Hamiltonian of the three orthogonal states (Fig. 2c). Thus, when the cost function is optimized to its global minima, the three output orthogonal states will be the three lowest eigenstates automatically. The used variational quantum circuit is shown in Fig. 2b. The circuit consists of multi-qubit $\sqrt{iSWAP}$ entangling gates with embedding variational single-qubit gates on both sides of them.

In our algorithmic experiment, the design parameter $E_L$, the inductive energy of the added Josephson junction array, is swept from 0.2 GHz to 3 GHz while the other parameters are fixed at $E_C = 0.7$ GHz, $E_J = 4.5$ GHz and $\phi_{ext} = \pi/2$. For each parameter, the cost function is optimized by the simultaneous perturbation stochastic approximation method[28]. Figure 2d shows the cost function and the corresponding energies of the three orthogonal states (the inset figure) as a function of the number of iterations in a typical design parameter at $E_L = 3.0$ GHz. We observe a competition phenomenon in the energies during the optimization: the energies of 1st and 2nd excited states first decrease and then increase to steady values, while the cost function decreases gradually. After completely sweeping the design parameter, we plot the energy spectrum of the three lowest eigenstates as a function of $E_L$ in Fig. 2e.

Then we used quantum error mitigation method to further reduce the decoherence error in the energy spectrum. Interestingly, in Fig. 2e, we see that the excited states have smaller average deviated energies (~0.52 GHz) than the ground state (~0.76 GHz). The reason is that, in contrast to the excited states, the quantum state error in the ground state can only increase the energy. These residual energy deviations are mainly contributed by the noise through the circuit decoherence, which can be reduced by quantum error mitigation method by considering the purities of the prepared quantum states (see Supplemental Material for more details). The results before and after error mitigation are represented by circles and crosses in Fig. 2e, respectively. We see that the average deviations of energy estimation have been reduced by 67%.

In the energy spectrum, we observe an anti-crossing signal between the two excited states when the inductive energy $E_L$ is near 0.45 GHz, which indicates the presence of a phase transition. The phase transition originates from the multiple valley profile of the potential in the system Hamiltonian. The physical picture is shown in Fig. 3a and 3b. Before the transition, the potential has two lowest valleys located around the phase 0 and $2\pi$ (Fig. 3a for $E_L = 0.2$ GHz) and the eigenstates $|0\rangle$ and $|1\rangle$ localized in different valleys. As the parameter $E_L$ increases, all of the potential valleys together with the eigenstates localized in them are lifted. When the energies of excited eigenstates $|1\rangle$ and $|2\rangle$ in different valleys become closer, there exists an anti-crossing in the energy spectrum. When the parameter $E_L$ further increases, there will be only one potential valley left in phase space, and the lowest two energy states are both located in this single valley (Fig. 3b for $E_L = 2.0$ GHz).

The left side of the anti-crossing is corresponding to the heavy Fluxonium qubit, which have attracted many attentions due to its long relaxation time. In contrast, at the right side, the two lowest-energy quantum states define a viable qubit driven by plasmon transition[15]. This is an interesting regime to implement advanced post-transmon qubits. We nickname it "Plasmonium" for the convenience of the following description.

Plasmonium is similar to the transmon qubit in terms of its control method, including the qubit initialization, manipulation, and implementation of inter-qubit coupling due to its moderate transition frequency, transition matrix elements, and charging energy (Extended Data Fig. 2). It is important to note that the undesirable trade-off between anharmonicity and charge noise sensitivity in the transmon has now been avoided by the added shunted inductance[14]. In Fig. 3c, we compare the flux sensitivity of Plasmonium with the Transmon, heavy Fluxonium, high-coherence Fluxonium, Quarton, and capacitively shunted flux qubit (CSFQ)[29] and observe that the Plasmonium has the smallest flux sensitivity. Thus, the Plasmonium has a large anharmonicity and a long coherent time on a wide tunable band of transition frequency. These unique properties of the Plasmonium will bring key advantages over the transmon qubit, including a shrinking physical size[30], reducing state leakage[31], and alleviating frequency crowding problems in multi-qubit devices[32].

Next, we move to the experimental characterization of the fabricated Plasmonium samples. In Fig. 4a, we show the scanning electron microscopy (SEM) image of Plasmonium, which has a length of 240 μm, only 40% of a typical Transmon qubit. The key element, the inductance of $E_L = 2.20$ GHz, is implemented by 37 large-area Al/Al-oxide/Al Josephson junctions connected in series[33,34]. The qubit is tuned in transition frequency by an inductively coupled transmission line (TL), driven by another capacitively coupled TL, and measured by probing a dispersively coupled resonator through a readout TL (Fig. 4a). The control and readout circuits are like Transmon and are fully compatible with Transmon's experimental electronics system.

We first confirm the large anharmonicity of Plasmonium. The transition frequences $f_{01}$ and $f_{12}$ at different flux biases are probed by microwave pulses of varying frequencies (4~6 GHz). The experimental spectrum is shown in Fig. 4b. From the data, we see the anharmonicity has a minimum of 490 MHz at the flux insensitive point $\phi_{ext}=0$, which is about 50% larger than typical Transmon qubits and can be further increased by reducing the capacitance of Plasmonium. We then confirm Plasmonium also exhibits high coherence time. At the flux insensitive point, we measured its relaxation time $T_1 = 32.5 \pm 1.2$ μs and dephasing time $T_2 = 16.3 \pm 0.9$ μs (Fig. 4d). This result is better than the best Transmon qubit fabricated on the same chip. In Fig. 4c, we present the measured coherence time at different flux detuning. The result shows that the qubit still keeps dephasing time above 2 μs when detuning from flux insensitive point as much as 400 MHz, which demonstrates a large high-$T_2$ frequency band. With recent progress in new superconducting materials and surface treatment method[35-37], we expect that the lifetime and coherence time of Plasmonium qubits can be improved to several hundreds of microseconds.

Finally, we characterize the performance of a universal quantum gate set. Two Plasmonium qubits are capacitively coupled via a frequency-tunable coupler. The performance of simultaneous single-qubit gates and two-qubit iSWAP-like gates are measured by cross-entropy benchmarking (XEB)[38,39]. In these quantum gate operations, the larger anharmonicity of Plasmonium can lead to a larger drive amplitude of single-qubit gates and coupling strength in two-qubit gates. Therefore, we can implement fast quantum gates with reduced leakage errors. We implement 20 ns single-qubit gates and 10 ns two-qubit iSWAP-like gate and obtained the XEB average fidelity to be 99.85(1)% and 99.86(1)% for single-qubit gates and 99.58(3)% for two-qubit gates in Fig. 4d. We emphasize that this is only the first batch of Plasmonium device we fabricated, where, however, the measured fidelities of a universal gate set have already exceeded the Transmon qubits used in the quantum-aided design (see Table S2).

In summary, we have demonstrated quantum-aided design on a superconducting quantum processor and we verified that the new Plasmonium has superior performance over the used Transmon qubit. Though the scale of the quantum circuit in this first step can be simulated on current classical computers, our work points to a new promising way to use intermediate-scale quantum computers to design advanced noise-protected qubits, which will be used to reduce the resource overhead of physical qubits towards a full fault-tolerant quantum computer.

**Figure Captions**

**Fig. 1| The concept of quantum-aided design. (a)** Quantum-aided design uses the existing quantum computing resource to accelerate the development of next-generation quantum computers. It plays the same role as electronic computer-aided design for classical computers, which has made great achievements in the past decades. **(b)** The outline of our work. We used practical variational quantum computing on a Transmon-based quantum processor to simulate a superconducting quantum electronic circuit. The circuit is a Josephson-junction-array shunted Transmon model. We identified, fabricated, and tested new Plasmonium qubits guided by the quantum simulation result and conformed a better performance than the used Transmons.

**Fig. 2| Variational quantum-aided design. (a)** Fock states up to 7 photons are used as the bases to encode the system Hamiltonian. **(b)** The variational quantum circuit is used to rotate the Fock states into the eigenstates. It has 26 variational parameters in the embedded single-qubit gates. The cost function is defined as a weighting sum of the three lowest energies: $E_0$, $E_1$ and $E_2$. **(c)** The system Hamiltonian is shown in the Fock bases. It is mapped to a weighting sum of Pauli operators (20 terms) through Gray code to estimate the energy eigenvalues. **(d)** A typical optimization process for design parameter $E_L = 3.0$ GHz. The inset shows the three lowest energy eigenvalues as a function of the number of iterations. **(e)** The energy spectrum of the three lowest eigenvalues as a function of the parameter $E_L$. The error-mitigated results are also shown.

**Fig. 3| Characteristics of the designed qubit. (a)(b)** Potential profile of the system Hamiltonian and the corresponding energy eigenvalues before ( $E_L = 0.2$ GHz ) and after ( $E_L = 2.0$ GHz ) the phase transition, respectively. **(c)** Flux noise sensitivity as a

function of the detuning from the flux insensitive point. Six types of qubits are put into comparison: Transmon, heavy Fluxonium, high-coherence Fluxonium, Quarton, capacitively shunted flux qubit (CSFQ), and Plasmonium.

**Fig. 4| Fabrication and test of the new qubit. (a)** The SEM images of the Plasmonium qubit. The geometry layout of the new qubit is similar to the Transmon qubit. The zoom-in shows the inductively shunted Josephson junction array. **(b)** The transition spectrum of the Plasmonium qubit is a function of the external flux. The $|0\rangle \leftrightarrow |1\rangle$ and $|1\rangle \leftrightarrow |2\rangle$ transitions are theoretically fitted by the green and red solid lines respectively. The additional gray dotted line comes from the non-ideal input of the microwave driving signal. **(c)** The dephasing time as a function of the detuning from its flux insensitive point. **(d)** The measurement of energy relaxation time and the dephasing time at the flux insensitive point. **(e)** The cross-entropy benchmarking of single and two-qubit gates. A single cycle in single-qubit gate benchmarking includes a single qubit gate of Q1 and Q2, and a single cycle in two-qubit gate benchmarking includes a layer of single-qubit gates and a layer of two-qubit gate.

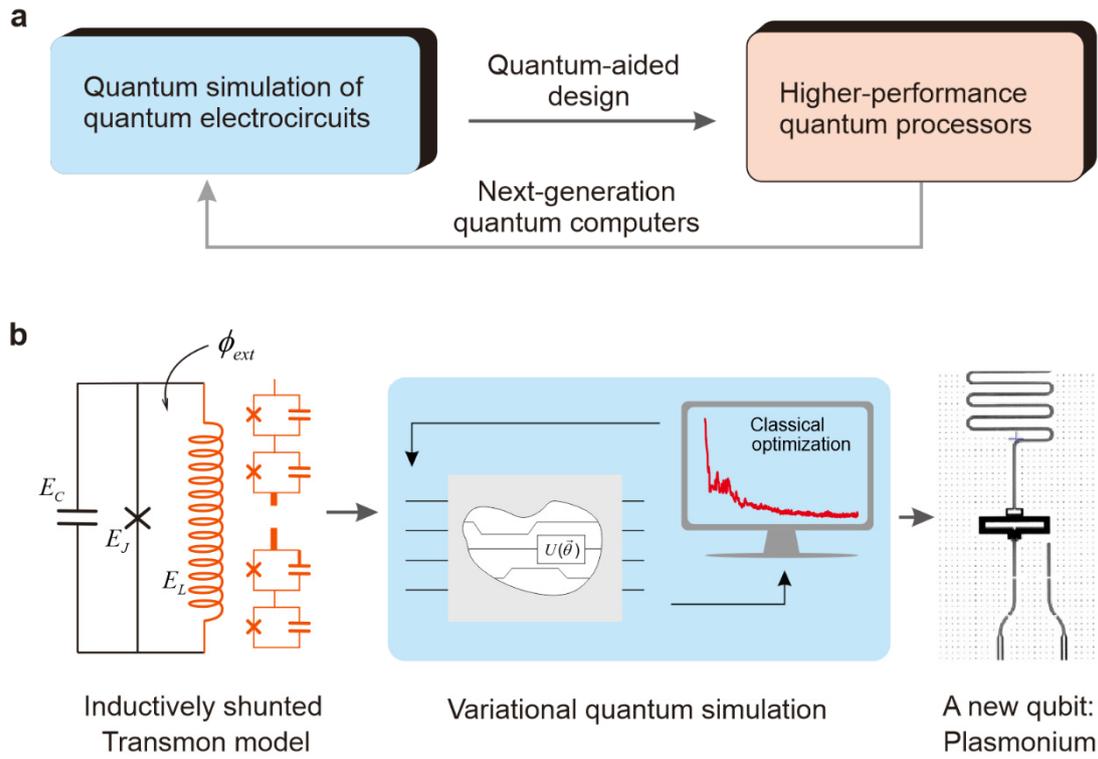

Figure 1

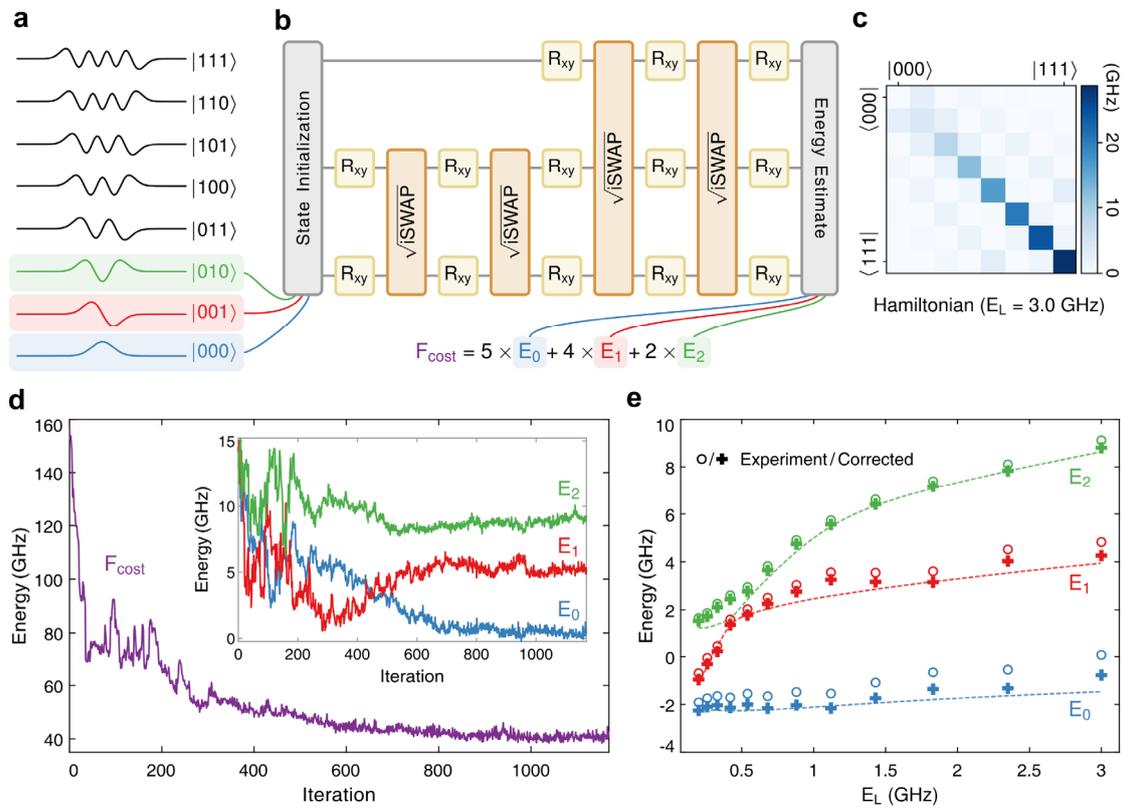

Figure 2

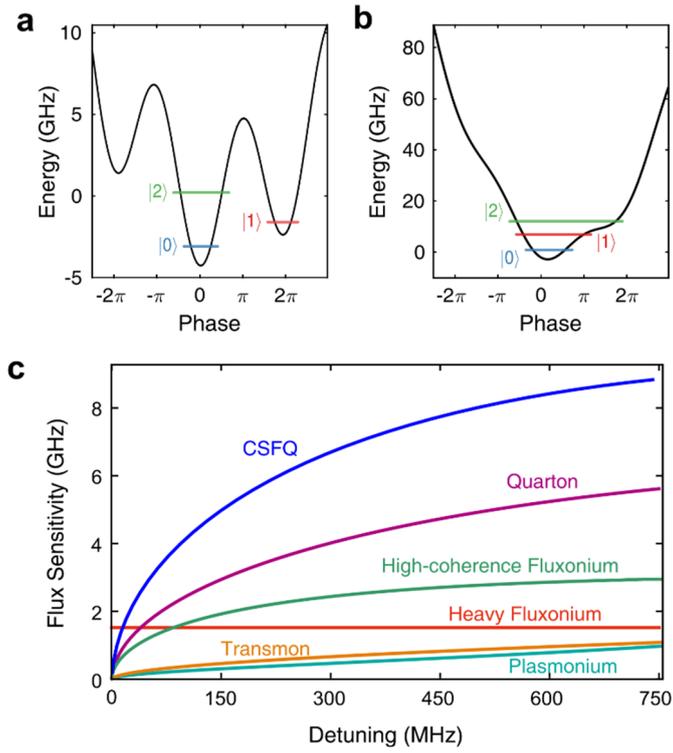

Figure 3

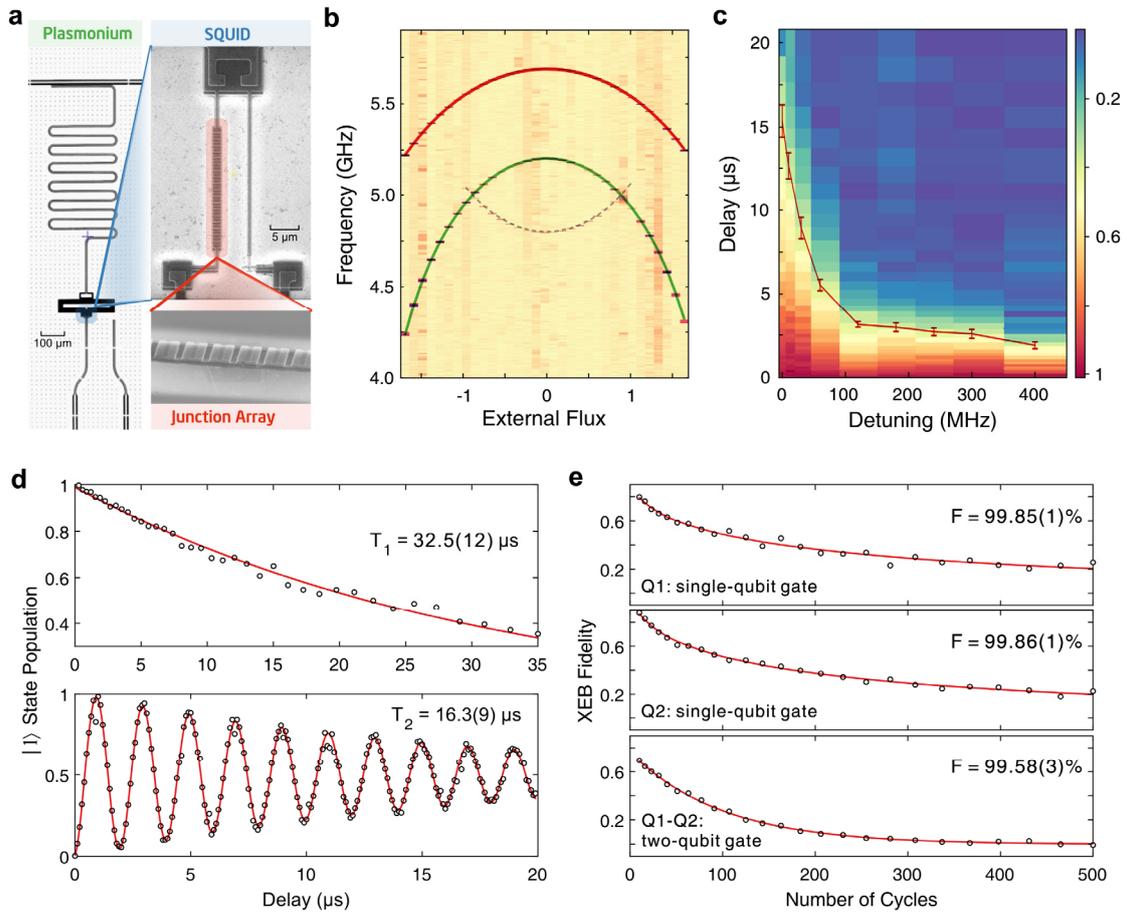

Figure 4

**Extended Data Fig. 1| Design parameters of energy scales and flux sweet spots of different types of superconducting qubits.** The X, Y, and Z axes are the ratios of $E_J/E_L$, $E_J/E_C$, and the positions of flux sweet spots, respectively. We chose a relatively large $E_L$ for the Plasmonium qubit. This is for obtaining a relatively larger frequency tunable range. Because we use the plasmon transition in our qubit, the coherence time does not decrease significantly with the increase of $E_L$. The $E_J/E_C$ of the Plasmonium qubit is smaller than the Transmon qubit but larger than the high-coherence Fluxonium and the quasi-charge qubit. This results in a larger anharmonicity while keeping the computational basis states localized in a single potential well. The Plasmonium qubit is operated around the sweet spot of zero flux quanta, which is the same as the Transmon qubit. In contrast, the heavy Fluxonium qubit operates at somewhere between zero and a half flux quanta without sweet spot, the high-coherence Fluxonium and Quarton are operated at half flux quanta, and the quasi-charge qubit has two sweet spots, both at zero and a half flux quanta.

**Extended Data Fig. 2| Comparison of operation parameters between different superconducting qubits. (a)** The transition frequency. **(b)** The transition matrix element. **(c)** The position of flux sweet spot. We observe that the Plasmonium qubit has a more moderate transition matrix element than heavy Fluxonium and a more moderate transition frequency than high-coherence Fluxonium. The former character contributes to the feasibility of single-qubit rotation and multi-qubit coupling while the latter makes the qubit easy to be initialized because of its smaller population in excited states at the thermal equilibrium state.

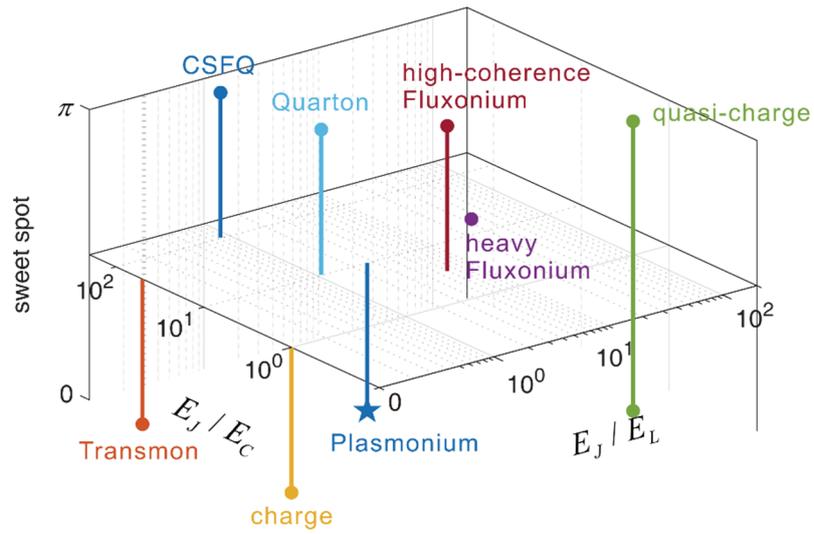

Extended Data Figure 1

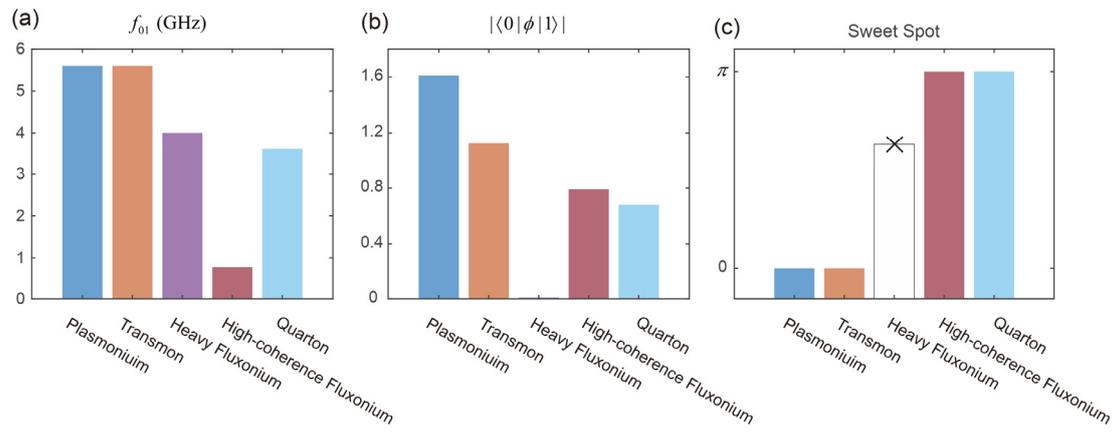

Extended Data Figure 2